\documentclass[aps,apl,twocolumn,groupedaddress,amsmath,reprint]{revtex4-1}
\usepackage[latin1]{inputenc}    
\usepackage{graphicx}   
\usepackage{xspace}
\usepackage{color}

\begin{document}

\newcommand{\mpar}[1]{\marginpar{\small \it \textcolor{magenta}{#1}}}

\newcommand*{\fmw}{\ensuremath{f_\mathrm{mw}}\xspace}
\newcommand*{\Ht}{\ensuremath{\mathcal{H}}\xspace}
\newcommand*{\Hi}{\ensuremath{\mathcal{H}^{-1}}\xspace}
\newcommand*{\Psw}{\ensuremath{P_\mathrm{sw}}\xspace}
\newcommand*{\tpw}{\ensuremath{t_\mathrm{pw}}\xspace}
\newcommand*{\rotq}{\ensuremath{\theta}\xspace}
\newcommand*{\Aq}{\ensuremath{A_\mathrm{Q}}\xspace}
\newcommand*{\Ai}{\ensuremath{A_\mathrm{I}}\xspace}
\newcommand*{\Hint}{\ensuremath{\hat{H}_\mathrm{int}}\xspace}
\newcommand*{\Hsub}{\ensuremath{\hat{H}_\mathrm{sub}}\xspace}
\newcommand*{\Hqb}{\ensuremath{\hat{H}_\mathrm{qb}}\xspace}
\newcommand*{\Htls}{\ensuremath{\hat{H}_\mathrm{TLS}}\xspace}
\newcommand*{\siX}{\ensuremath{\hat{\sigma}_\mathrm{x}}\xspace}
\newcommand*{\siY}{\ensuremath{\hat{\sigma}_\mathrm{y}}\xspace}
\newcommand*{\siZ}{\ensuremath{\hat{\sigma}_\mathrm{z}}\xspace}
\newcommand*{\siXq}{\ensuremath{\hat{\sigma}_\mathrm{x}^\mathrm{qb}}\xspace}
\newcommand*{\siYq}{\ensuremath{\hat{\sigma}_\mathrm{y}^\mathrm{qb}}\xspace}
\newcommand*{\siZq}{\ensuremath{\hat{\sigma}_\mathrm{z}^\mathrm{qb}}\xspace}
\newcommand*{\siXt}{\ensuremath{\hat{\sigma}_\mathrm{x}^\mathrm{TLS}}\xspace}
\newcommand*{\siYt}{\ensuremath{\hat{\sigma}_\mathrm{y}^\mathrm{TLS}}\xspace}
\newcommand*{\siZt}{\ensuremath{\hat{\sigma}_\mathrm{z}^\mathrm{TLS}}\xspace}
\newcommand*{\siXs}{\ensuremath{\hat{\sigma}_\mathrm{x}^\mathrm{sub}}\xspace}
\newcommand*{\siYs}{\ensuremath{\hat{\sigma}_\mathrm{y}^\mathrm{sub}}\xspace}
\newcommand*{\siZs}{\ensuremath{\hat{\sigma}_\mathrm{z}^\mathrm{sub}}\xspace}
\newcommand*{\Isq}{\ensuremath{I_\mathrm{b}}\xspace}
\newcommand*{\PhiQ}{\ensuremath{\Phi_\mathrm{qb}}\xspace}
\newcommand*{\fqb}{\ensuremath{\omega_\mathrm{qb}}\xspace}
\newcommand*{\ftls}{\ensuremath{f_\mathrm{TLS}}\xspace}
\newcommand*{\fosc}{\ensuremath{f_\mathrm{osc}}\xspace}
\newcommand*{\dph}{\ensuremath{\delta\Phi}\xspace}
\newcommand*{\df}{\ensuremath{\delta\!f}\xspace}
\newcommand*{\dpint}{\ensuremath{\delta\Phi_\mathrm{int}}\xspace}
\newcommand*{\TphN}{\ensuremath{T_{\varphi,\mathrm{N}}}\xspace}
\newcommand*{\Tph}[1]{\ensuremath{T_{\varphi,\mathrm{#1}}}\xspace}
\newcommand*{\tp}{\ensuremath{t_\mathrm{p}}\xspace}
\newcommand*{\fRabi}{\ensuremath{f_\mathrm{Rabi}}\xspace}
\newcommand*{\Ian}{\ensuremath{I_\mathrm{antenna}^\mathrm{mw}}\xspace}
\newcommand*{\Imr}{\ensuremath{I_\mathrm{r}^\mathrm{mw}}\xspace}
\newcommand*{\emd}{\ensuremath{\varepsilon^\mathrm{mw}_\mathrm{direct}}\xspace}
\newcommand*{\emt}{\ensuremath{\varepsilon^\mathrm{mw}}\xspace}

\newcommand{\ket}[1]{\vert  #1 \rangle} 
\newcommand{\bra}[1]{\langle  #1 \vert}

\newcommand*{\PhiX}{\ensuremath{\Phi_\mathrm{X}}\xspace}
\newcommand*{\PhiZ}{\ensuremath{\Phi_\mathrm{Z}}\xspace}
\newcommand*{\fX}{\ensuremath{f_\mathrm{x}}\xspace}
\newcommand*{\fZ}{\ensuremath{f_\mathrm{z}}\xspace}
\newcommand*{\Ax}{\ensuremath{A_\mathrm{x}}\xspace}
\newcommand*{\Az}{\ensuremath{A_\mathrm{z}}\xspace}

\newcommand*{\TF}{\ensuremath{T_{\varphi F}}\xspace}
\newcommand*{\TE}{\ensuremath{T_{\varphi E}}\xspace}
\newcommand*{\GF}{\ensuremath{\Gamma_{\varphi F}}\xspace}
\newcommand*{\GE}{\ensuremath{\Gamma_{\varphi E}}\xspace}

\newcommand*{\GSs}{\ensuremath{\,\mathrm{GS/s}\xspace}}
\newcommand*{\mPh}{\ensuremath{\,\mathrm{m}\Phi_0}\xspace}
\newcommand*{\uPh}{\ensuremath{\,\mu\Phi_0}\xspace}

\newcommand*{\um}{\ensuremath{\,\mu\mathrm{m}}\xspace}
\newcommand*{\nm}{\ensuremath{\,\mathrm{nm}}\xspace}
\newcommand*{\mm}{\ensuremath{\,\mathrm{mm}}\xspace}
\newcommand*{\m}{\ensuremath{\,\mathrm{m}}\xspace}
\newcommand*{\sqm}{\ensuremath{\,\mathrm{m}^2}\xspace}
\newcommand*{\sqmm}{\ensuremath{\,\mathrm{mm}^2}\xspace}
\newcommand*{\squm}{\ensuremath{\,\mu\mathrm{m}^2}\xspace}
\newcommand*{\psqm}{\ensuremath{\,\mathrm{m}^{-2}}\xspace}
\newcommand*{\psqmV}{\ensuremath{\,\mathrm{m}^{-2}\mathrm{V}^{-1}}\xspace}
\newcommand*{\cm}{\ensuremath{\,\mathrm{cm}}\xspace}

\newcommand*{\nF}{\ensuremath{\,\mathrm{nF}}\xspace}
\newcommand*{\pF}{\ensuremath{\,\mathrm{pF}}\xspace}
\newcommand*{\pH}{\ensuremath{\,\mathrm{pH}}\xspace}

\newcommand*{\emob}{\ensuremath{\,\mathrm{m}^2/\mathrm{V}\mathrm{s}}\xspace}
\newcommand*{\edos}{\ensuremath{\,\mu\mathrm{C}/\mathrm{cm}^2}\xspace}
\newcommand*{\mbar}{\ensuremath{\,\mathrm{mbar}}\xspace}

\newcommand*{\A}{\ensuremath{\,\mathrm{A}}\xspace}
\newcommand*{\mA}{\ensuremath{\,\mathrm{mA}}\xspace}
\newcommand*{\nA}{\ensuremath{\,\mathrm{nA}}\xspace}
\newcommand*{\pA}{\ensuremath{\,\mathrm{pA}}\xspace}
\newcommand*{\fA}{\ensuremath{\,\mathrm{fA}}\xspace}
\newcommand*{\uA}{\ensuremath{\,\mu\mathrm{A}}\xspace}

\newcommand*{\Ohm}{\ensuremath{\,\Omega}\xspace}
\newcommand*{\kOhm}{\ensuremath{\,\mathrm{k}\Omega}\xspace}
\newcommand*{\MOhm}{\ensuremath{\,\mathrm{M}\Omega}\xspace}
\newcommand*{\GOhm}{\ensuremath{\,\mathrm{G}\Omega}\xspace}

\newcommand*{\Hz}{\ensuremath{\,\mathrm{Hz}}\xspace}
\newcommand*{\kHz}{\ensuremath{\,\mathrm{kHz}}\xspace}
\newcommand*{\MHz}{\ensuremath{\,\mathrm{MHz}}\xspace}
\newcommand*{\GHz}{\ensuremath{\,\mathrm{GHz}}\xspace}
\newcommand*{\THz}{\ensuremath{\,\mathrm{THz}}\xspace}

\newcommand*{\K}{\ensuremath{\,\mathrm{K}}\xspace}
\newcommand*{\mK}{\ensuremath{\,\mathrm{mK}}\xspace}

\newcommand*{\kV}{\ensuremath{\,\mathrm{kV}}\xspace}
\newcommand*{\V}{\ensuremath{\,\mathrm{V}}\xspace}
\newcommand*{\mV}{\ensuremath{\,\mathrm{mV}}\xspace}
\newcommand*{\uV}{\ensuremath{\,\mu\mathrm{V}}\xspace}
\newcommand*{\nV}{\ensuremath{\,\mathrm{nV}}\xspace}

\newcommand*{\eV}{\ensuremath{\,\mathrm{eV}}\xspace}
\newcommand*{\meV}{\ensuremath{\,\mathrm{meV}}\xspace}
\newcommand*{\ueV}{\ensuremath{\,\mu\mathrm{eV}}\xspace}

\newcommand*{\T}{\ensuremath{\,\mathrm{T}}\xspace}
\newcommand*{\mT}{\ensuremath{\,\mathrm{mT}}\xspace}
\newcommand*{\uT}{\ensuremath{\,\mu\mathrm{T}}\xspace}

\newcommand*{\ms}{\ensuremath{\,\mathrm{ms}}\xspace}
\newcommand*{\s}{\ensuremath{\,\mathrm{s}}\xspace}
\newcommand*{\us}{\ensuremath{\,\mathrm{\mu s}}\xspace}
\newcommand*{\ns}{\ensuremath{\,\mathrm{ns}}\xspace}
\newcommand*{\rpm}{\ensuremath{\,\mathrm{rpm}}\xspace}
\newcommand*{\minute}{\ensuremath{\,\mathrm{min}}\xspace}
\newcommand*{\degree}{\ensuremath{\,^\circ\mathrm{C}}\xspace}

\newcommand*{\EqRef}[1]{Eq.\,(\ref{#1})}
\newcommand*{\FigRef}[1]{Fig.\,\ref{#1}}
\newcommand*{\dd}[2]{\mathrm{\partial}#1/\mathrm{\partial}#2}
\newcommand*{\ddf}[2]{\frac{\mathrm{\partial}#1}{\mathrm{\partial}#2}}

\title{Improving quantum gate fidelities by using a qubit\\ to measure microwave pulse distortions}
 \author{Simon Gustavsson$^1$}
 \author{Olger Zwier$^{1,\dag}$}
 \author{Jonas Bylander$^{1}$}
 \author{Fei Yan$^{2}$}
 \author{Fumiki Yoshihara$^3$}
 \author{Yasunobu Nakamura$^{3,4}$}
 \author{Terry P. Orlando$^{1}$}
 \author{William D. Oliver$^{1,5}$}
 \affiliation{$^1$Research Laboratory of Electronics, Massachusetts Institute of Technology, Cambridge, MA 02139, USA \\
  $^2$Department of Nuclear Science and Engineering, MIT, Cambridge, MA 02139, USA \\
  $^3$The Institute of Physical and Chemical Research (RIKEN), Wako, Saitama 351-0198, Japan \\
  $^4$Research Center for Advanced Science and Technology (RCAST), The University of Tokyo, Komaba, Meguro-ku, Tokyo 153-8904, Japan\\
  $^5$MIT Lincoln Laboratory, 244 Wood Street, Lexington, MA 02420, USA \\
  $^\dag$Present address:  Zernike Institute for Advanced Materials, University of Groningen, 9747AG Groningen, The Netherlands}

\begin{abstract}
We present a new method for determining pulse imperfections and improving the single-gate fidelity in a superconducting qubit.  
By applying consecutive positive and negative $\pi$ pulses, we amplify the qubit evolution due to microwave pulse distortion, which causes the qubit state to rotate around an axis perpendicular to the intended rotation axis.
Measuring these rotations as a function of pulse period allows us to reconstruct the shape of the microwave pulse arriving at the sample.
Using the extracted response to predistort the input signal, we are able to improve the pulse shapes and to reach an average single-qubit gate fidelity higher than $99.8\%$. 
\end{abstract}

\maketitle
A basic requirement for building a quantum information processor is the ability to perform fast and precise single- and two-qubit gate operations \cite{Bremner:2002}.
For qubits defined in superconducting circuits, much work has been done to improve the quality of both single-qubit \cite{Lucero:2008, Chow:2009, Chow:2010} and two-qubit gate operations \cite{Kerman:2008,Bialczak:2010,Chow:2011,Dewes:2012, GustavssonTLS:2012,Steffen:2012,Chow:2012}. 
Still, gate fidelities need to improve further to reach error rates small enough for practically implementing fault-tolerant quantum computing with error-correcting protocols \cite{Knill:2005, DiVincenzo:2009}.
In most qubit architectures, many single-qubit operations are implemented by applying short microwave pulses resonant with the qubit transition frequency.  The phase of the microwave pulse controls the rotation axis in the \emph{x-y} plane of the Bloch sphere, whereas the pulse amplitude and duration set the rotation angle.
A difficulty with this approach is that the single-qubit gate fidelity becomes highly susceptible to any impedance mismatch in the microwave line between the signal generator and the qubit, since such imperfections lead to pulse distortions. 

\begin{figure}[t!]
\centering
\includegraphics[width=\linewidth]{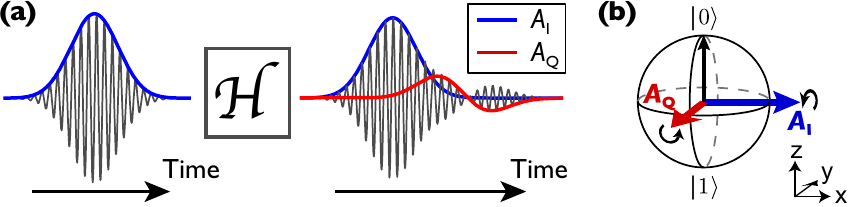}
\caption{(a) Distortion of a Gaussian-shaped microwave pulse.  When passing through the transfer function $\mathcal{H}$, the pulse shape gets distorted and acquires a quadrature component $\Aq$.  
(b) Bloch sphere describing the qubit dynamics in the rotation frame.  The drive fields $\Ai$ and $\Aq$ generate rotations around the $x$- and $y$-axes, respectively.  The distorted pulse shape in (a) will cause rotations around an axis non-parallel to $x$, thereby generating errors in the final qubit state.
}
\label{fig:Sketch}
\end{figure}

Consider the microwave pulse shown in \FigRef{fig:Sketch}(a), which initially has a Gaussian-shaped envelope $\Ai(t)$ with a well-defined phase.  
When passing from the generator to the device, the pulse is distorted and acquires a quadrature component $\Aq(t)$. 
The pulse was intended to perform a rotation around the \emph{x}-axis of the Bloch sphere [see \FigRef{fig:Sketch}(b)], but the quadrature components present in the distorted pulse shape will change the rotation axis and generate an error in the final qubit state.  The systematic errors due to the non-zero $\Aq(t)$ are particularly problematic for qubit control, since they will bring the qubit state out of the \emph{y-z} plane expected from a pure rotation around the \emph{x}-axis.

The distortion is described by the transfer function $\mathcal{H}$, which is the frequency-domain representation of the system's impulse response $h(t)$.
%
If the transfer function is known, it is possible to correct pulse imperfections using digital signal processing techniques. By numerically applying the inverse $\mathcal{H}^{-1}$ to the input signal $x$, the pulse is predistorted in precisely the right way to give the correct signal $\mathcal{H}\left[\mathcal{H}^{-1}\left[x\right]\right] = x$ at the device.
The difficulty lies in obtaining $\mathcal{H}$. Since superconducting qubits operate at millikelvin temperatures inside a dilution refrigerator, it is generally not possible to probe the signal arriving at the qubit directly with conventional instruments such as a network analyzer or a sampling oscilloscope.


In this work, we take a different approach and use the qubit's response to various pulses as a probe for determining $\mathcal{H}$ \cite{Bylander:2009}. 
We have designed and implemented a pulse sequence aimed at obtaining the unwanted quadrature component of the signal arriving at the qubit.  
The sequence consists of pairs of positive and negative $\pi$ pulses around the \emph{x}-axis; the reversing of the pulse direction amplifies the quadrature component of the signal and causes the qubit to slowly oscillate around the \emph{y}-axis.
By measuring the rotation frequency for different pulse periods, we are able to extract the time dependence of those quadrature components.
From the obtained signal we construct the inverse transfer function $\mathcal{H}^{-1}$, and use it to numerically predistort the input signal. The resulting pulse shapes give a significant reduction in the gate error rate, as determined in a randomized benchmarking experiment \cite{Knill:2008}.  With optimized pulse shapes, we extract an average gate fidelity higher than $99.8\%$, which, to our knowledge, is the highest gate fidelity reported so far for a superconducting qubit.

We use a flux qubit \cite{Mooij:1999}, consisting of a superconducting loop interrupted by four Josephson junctions.  Biased at the optimal operation point, the qubit's energy relaxation time is $T_1=12\us$, and the dephasing time is $T_2^*=2.5\us$ (see Ref.\,\cite{Bylander:2011} for a detailed device description).  
The device is embedded in a SQUID, which is used as a sensitive magnetometer for qubit read-out \cite{Chiorescu:2003}.  We implement the read-out by applying a short current pulse to the SQUID to determine its switching probability $\Psw$.
When statically biasing the qubit loop at half a flux quantum $\Phi_0/2$ ($\Phi_0= h/2e$), the Hamiltonian becomes $H =  - \frac{\hbar}{2} \left( \fqb \, \siZ + A(t) \, \siX  \right)$,
where $\fqb /2\pi= 5.4\GHz$ is the qubit frequency and $A(t) = \Ai(t)\cos(\omega t) + \Aq(t) \sin(\omega t)$ is the drive field. The drive is generated by applying an oscillatory flux $\Phi(t)$ to the qubit loop using an on-chip antenna, giving $A(t) = 2 I_\mathrm{P} \Phi(t)/\hbar$, with $I_\mathrm{P}=180\nA$ being the loop's persistent current.  When driving the qubit resonantly ($\omega=\fqb$) and going to the rotating frame, we get 
\begin{equation}
  H =  - \frac{\hbar}{2} \left( \Ai(t) \, \siX^\mathrm{rot.} + \Aq(t) \, \siY^\mathrm{rot.} \right), \label{eq:Hr} 
\end{equation}
which is the Hamiltonian depicted in the Bloch sphere in \FigRef{fig:Sketch}(b).

The microwave pulses are created by generating in-phase [$\Ai(t)$] and quadrature [$\Aq(t)$] pulse envelopes using a Tektronix 5014 arbitrary waveform generator (AWG), and sending them to the internal IQ mixer of an Agilent 8267D microwave generator.  
We write the total transfer function from generator to qubit as $\mathcal{H} = \mathcal{H}_\mathrm{ext} \,\mathcal{H}_\mathrm{int}$, where $\mathcal{H}_\mathrm{ext}$ refers to imperfections in the electronics and coaxial cables outside the cryostat, and $\mathcal{H}_\mathrm{int}$ describe signal distortion occurring inside the cryostat, for example from bonding wires or impedance mismatches on the chip.
To ensure that the pulses we send to the cryostat are initially free from distortion, we determine $\mathcal{H}_\mathrm{ext}$ with a high-speed oscilloscope, and use $ \mathcal{H}_\mathrm{ext}^{-1}$ to correct for imperfections in the AWG and in the IQ mixers \cite{Hofheinz:2009,JohnsonThesis:2011}.  The setup allows us to create well-defined Gaussian-shaped microwave pulses with pulse widths as short as $\tpw=2.5\ns$ \footnote{See online supplementary material S1.}.  We define the Gaussian as $A(t) = A\exp[-\pi t^2/\tpw^2]$, so that the integrated area under the pulse (corresponding to the total rotation angle) equals $A\,\tpw$.  The pulses are truncated at a total duration of $3\tpw$.

\begin{figure}[tb]
\centering
\includegraphics[width=\linewidth]{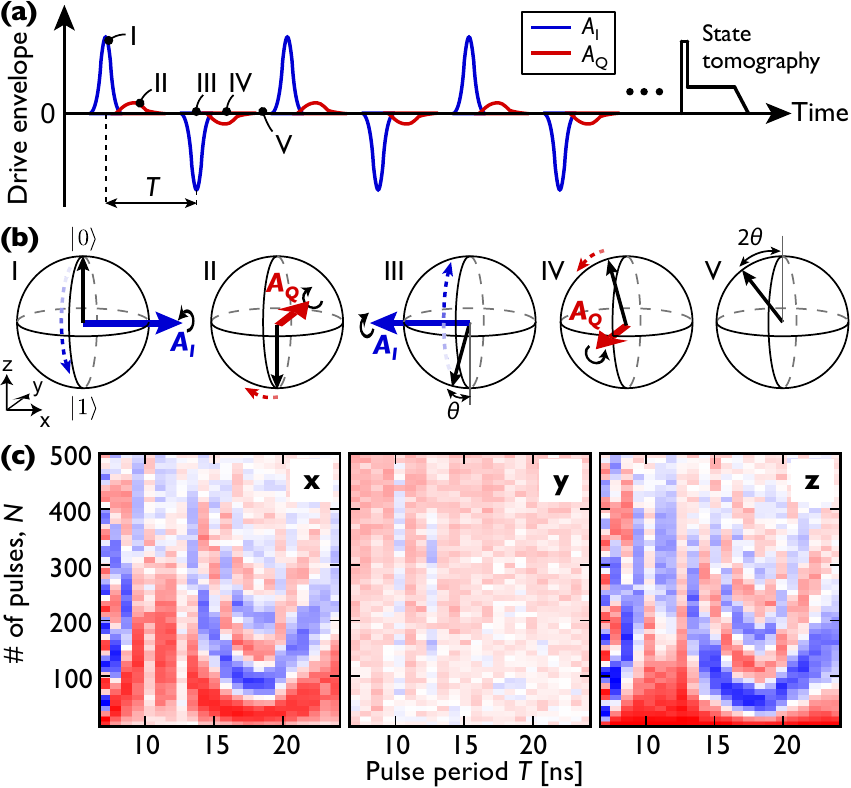}
\caption{(a) Pulse sequence used to probe quadrature components in the microwave pulses.  By applying consecutive pairs of pulses with positive/negative amplitude to $\Ai$, the qubit rotations due to the in-phase signal cancel out, leaving only contributions from the quadrature component $\Aq$.
(b) Bloch spheres depicting the evolution of the qubit during the pulse sequence in (a).  The angle $\rotq$ is the quadrature rotation acquired per $\pi$ pulse.
(c) Qubit population after the pulse sequence in (a), measured vs pulse period $T$ and total number of pulses $N$, and projected onto the three axis $x$, $y$ and $z$.  The qubit undergoes slow rotations in the \emph{x-z} plane due to the quadrature components in the microwave pulses. 
}
\label{fig:RotQ}
\end{figure}

To extract information about $\mathcal{H}_\mathrm{int}$, we drive the qubit with consecutive pairs of positive and negative $\pi$ pulses in $\Ai(t)$, separated by the pulse period $T$.  The sequence is depicted in \FigRef{fig:RotQ}(a), together with Bloch spheres describing the qubit states at various points of the pulse sequence.  Note that in \FigRef{fig:RotQ}(a), we show an example of the drive pulses that reach the qubit, including a small $\Aq$-distortion after each pulse to better illustrate how the sequence works.  The signal we create at the generator does not have any quadrature components.
Starting with the qubit in the ground state, we apply a $\pi$-pulse around $x$ to take the qubit to $\ket{1}$ [step I in \FigRef{fig:RotQ}(a-b)].  Next, the $\Aq$-part, due to the pulse distortion, induces a small rotation $\theta$ around $y$, bringing the qubit state slightly off the south pole (II).  The negative $\pi$ pulse then takes the qubit back towards the north pole (III), but since this pulse is inverted, the following $\Aq$-part rotates the state even further away from $\ket{0}$ (IV).  After the first two pulses, the qubit has acquired a rotation of $2\theta$ around the $y$-axis (V).  The sequence is then repeated, and for each pair of subsequent $\pi$ pulses the qubit rotates another $2\theta$. 

Figure \ref{fig:RotQ}(c) shows the qubit state after the pulse sequence, measured versus the number of pulses and the pulse period $T$, and projected onto the three axes $x$, $y$, and $z$ using additional $\pi/2$ pulses to do state tomography before reading out the qubit's polarization \cite{Steffen:2006}.  There are clear oscillations in the $x$- and $z$-components, verifying that the qubit indeed rotates around the $y$-axis despite the pulses being applied to $x$.  Note that the rotation frequency is relatively slow: it typically takes a few hundred $\pi$ pulses to perform one full rotation around $y$.
A striking feature of \FigRef{fig:RotQ}(c) is that the oscillation frequency varies with pulse period all the way up to $T=25\ns$, much longer than the pulse width $\tpw=2.5\ns$.  This indicates that the quadrature distortions persist for a substantial time after the pulse should have ended. 

\begin{figure}[tb]
\centering
\includegraphics[width=\linewidth]{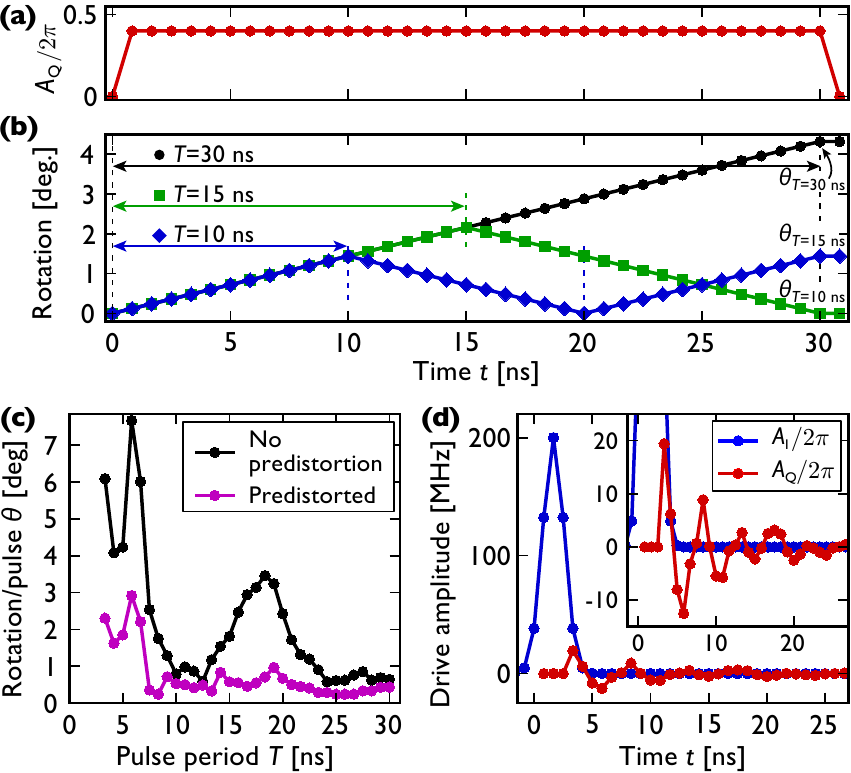}
\caption{(a) Constant quadrature distortion $\Aq(t)$ used to illustrate how the rotation angle depends on pulse period. 
(b) Qubit quadrature rotation for the quadrature distortion shown in (a), calculated with the pulse sequence from \FigRef{fig:RotQ}(a) and plotted for a few different pulse periods $T$.
(c) Quadrature rotation $\rotq$ acquired per $\pi$ pulse.  The black points are extracted from data similar to the measurement shown in \FigRef{fig:RotQ}(c). The magenta points are the results using a pre-distorted pulse shape, aimed at minimizing the quadrature distortion.
(d) Quadrature component $\Aq$ appearing at the sample when applying a $2.5\ns$ wide Gaussian pulse $\Ai$ at the input of the experimental setup.  The signal $\Aq$ is extracted from the quadrature rotations shown in (c) (black data points).
}
\label{fig:Response}
\end{figure}

To explain why the quadrature rotations depend on pulse period, we need to understand what happens when the $\pi$ pulses start to overlap with the distortions of the previous $\pi$ pulses.  Let us start by assuming that the $\pi$ pulses are instantaneous, and consider the qubit response to the static quadrature distortion shown in \FigRef{fig:Response}(a), where $\Aq(t)$ remains constant at $\Aq/2\pi=0.4\MHz$ for $30\ns$ after the $\pi$-pulse in $\Ai(t)$ at $t=0$.  
Figure \ref{fig:Response}(b) shows the qubit quadrature rotation during the distortion, plotted for different values of the pulse period $T$.  If $T$ is $30\ns$ or longer [black circles in \FigRef{fig:Response}(b)], the qubit will continuously rotate in one direction during $\Aq(t)$, acquiring a total rotation per pulse of $\theta = \int_{t=0}^{t=30\ns} \Aq(t)dt \approx 4.3 \deg$.  
However, if the pulse period is only $T=15\ns$ [green squares in \FigRef{fig:Response}(b)], the second $\pi$ pulse in $\Ai$ at $t=15\ns$ will reverse the direction of the $\Aq$-induced rotations of the first pulse, in the same way that a $\pi$ pulse in a spin-echo experiment reverses the spin evolution due to low-frequency field fluctuations in its environment \cite{Hahn:1950}. The rotation per pulse $\theta$ acquired with pulse period $T=15\ns$ ends up being zero, since the rotations during the second half of $\Aq(t)$ exactly cancel out the rotations during the first half.  For $T=10\ns$ [blue diamonds in \FigRef{fig:Response}(b)], there are two extra $\pi$ pulses in $\Ai$ occurring during the distortions of the first pulse, and we end up with $\theta=1.4\deg$. 
%
Note that we only consider the rotation due to the distortion of the first $\pi$ pulse; the total qubit rotation will be a sum of the rotations from all pulses. 

%
Having understood why $\theta$ depends on pulse period $T$ for a given $\Aq(t)$, we now ask if we can invert the problem:  given a measurement of $\theta$ as a function of $T$ such as the black trace in \FigRef{fig:Response}(c), can we extract the pulse distortions $\Aq(t)$?
To simplify the problem, we discretize time in the smallest steps available with our AWG, $\Delta t = 1/(1.2\GSs) \approx 0.83\ns$, and write $\Aq(t)$ as a vector $\vec{Q} = [Q_1,Q_2,\cdots, Q_{N}]$, with $Q_n =\Aq(n\Delta t)$.  The rotations $\theta(T)$ in \FigRef{fig:Response}(a) are measured with the same time resolution, and in a similar fashion we write $\theta(T)$ as $\vec{\theta} = [\theta_1,\theta_2,\cdots,\theta_{N}]$, $\theta_m =\theta(m\Delta t)$.
Both vectors contain $N = 30\ns/\Delta t = 36$ elements.
%
We still assume the $\pi$ pulses in $\Ai$ to be instantaneous, occurring with a period of $m=T/\Delta t$ in the discretized time.  

As explained previously, the $\pi$ pulses will act to periodically reverse the direction of the $\vec{Q}$-induced rotations, and the total rotation angle $\theta_m$ generated by $\vec{Q}$ becomes a sum of forward and backward rotations, depending on the period of the $\pi$ pulses:
\begin{equation}
\theta_m = \Delta t \left[\sum_{n=1}^m Q_n - \sum_{n=m+1}^{2m} Q_n +\sum_{n=2m+1}^{3m} Q_n - \cdots \right]
\label{eq:R1}  
\end{equation}
We can write \EqRef{eq:R1} as a system of linear equations $\vec{\theta} = \Delta t \,\mathbf{M} \, \vec{Q}$,
where $\mathbf{M}$ is a matrix with elements being either $1$ or $-1$ depending on the direction of rotation \footnote{See online supplementary material S2.}.  By inverting the matrix, we get the quadrature distortions directly from the measured rotations $\theta$:
\begin{equation}
\vec{Q} = \mathbf{M}^{-1} \, \vec{\theta} / \Delta t.
\label{eq:Res}  
\end{equation}

In the experiment, the $\pi$ pulses have a finite width $\tpw = 2.5\ns$.
During the pulses, the qubit is strongly driven around the $x$-axis, and the quadrature rotations due to $\Aq$ are suppressed to first order in $\Aq/\Ai$ during those $\tpw/\Delta t = 3$ discrete time steps when $\Ai\gg \Aq$.   The vectors therefore have to be limited to $\vec{Q} = [Q_4,Q_5,\cdots, Q_{N}]$ and $\vec{\theta} = [\theta_4,\theta_5,\cdots, \theta_{N}]$, and the matrix $\mathbf{M}$ needs to be modified to include zeros at the positions of the $\pi$ pulses in $\Ai$ \footnote{See online supplementary material S3.}.

In \FigRef{fig:Response}(d), we show the extracted quadrature response $\vec{Q}$, calculated using \EqRef{eq:Res} and the rotation data $\theta$ from \FigRef{fig:Response}(c).  For reference we also plot the shape and amplitude of the intended drive pulse $\Ai$, digitized at $I_n =\Ai(n\Delta t)$.  The pulse has an amplitude of $200\MHz$, giving a $\pi$ rotation in $\tpw=2.5\ns$.
The extracted quadrature response $\vec{Q}$ has considerably lower amplitude, but keeps oscillating for $25\ns$ after the main pulse ends.

Next, we use the measured response shown in \FigRef{fig:Response}(d) to determine the transfer function $\mathcal{H}_\mathrm{int}$ of the system \footnote{See online supplementary material S4.}.  With knowledge of $\mathcal{H}$, we can calculate the inverse $\mathcal{H}^{-1}$ and use it to predistort the input signal, with the aim of reducing the quadrature distortions.  The magenta trace in \FigRef{fig:Response}(c) shows the quadrature rotations $\theta$ for the same sequence of positive/negative pulses, but this time measured with a predistorted input signal.  Compared to the black trace, $\theta$ has been significantly reduced for all values of the pulse period $T$, thus validating our method and confirming that the pulse shown in \FigRef{fig:Response}(d) actually corresponds to the signal appearing at the sample.
We attribute the rotations still present after predistortion to errors due to 
oversimplifications in the linear model in \EqRef{eq:Res} used to extract $\vec{Q}$.  It may be possible to get a better estimate for $\Aq$ by calculating the qubit response using the full dynamics of the Hamiltonian in \EqRef{eq:Hr}, but it would involve solving a system of 36 non-linear equations, which computationally is much harder than inverting the matrix $\mathbf{M}$ in \EqRef{eq:Res}. 

Note that there may also be pulse distortions appearing in the in-phase component $\Ai$.  However, the consecutive positive and negative pulses in the sequence of \FigRef{fig:RotQ}(a) will cancel the effect of any errors in the rotations around $x$, which is also confirmed in the experiment (the $y$-component in \FigRef{fig:RotQ}(c) shows no oscillations).  
This cancellation allows us to exclusively target the $\Aq$-distortions.


\begin{figure}[tb]
\centering
\includegraphics[width=\linewidth]{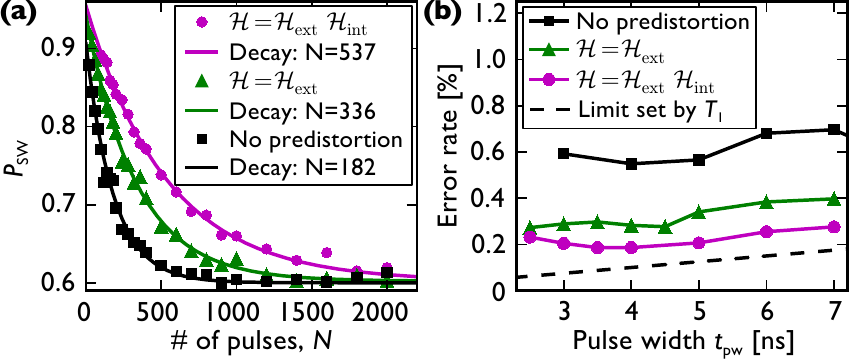}
\caption{
(a) Qubit polarization as a function of the number of pulses in the RBM sequence, measured with and without predistortion. The pulses have width $\tpw=3.5\ns$, separated by the period $T= 10.5\ns$. We use the same number of sequences and randomizations as in Refs.\,\cite{Knill:2008,Chow:2009}.
(b) Average error per pulse versus pulse width $\tpw$, measured with $T=3\,\tpw$.  The predistorted pulses give a lower error rate for all pulse widths.
}
\label{fig:RBM}
\end{figure}

Having determined a way to reduce quadrature distortions and improve the microwave pulse shapes, we proceed to characterize the qubit gate fidelity in our system.  A convenient way of testing single-qubit gates is to implement the randomized benchmarking protocol (RBM) \cite{Knill:2008}, where a random sequence of $\pi$ and $\pi/2$ pulses around the $x$- and $y$-axes are applied to the qubit.  If the pulses are imperfect, the qubit will start to dephase as the pulse errors accumulate.
%
Figure~\ref{fig:RBM}(a) shows examples of decay traces, where the three traces correspond to data measured with either full predistortion ($\mathcal{H} = \mathcal{H}_\mathrm{ext} \,\mathcal{H}_\mathrm{int}$), with predistortion only for the room-temperature electronics ($\mathcal{H} = \mathcal{H}_\mathrm{ext}$), or with no predistortion at all.  The pulses with full predistortion give a significantly slower decay than those without; when fitting to an exponential decay we find a decay constant of $N=537\pm22$ pulses, corresponding to an average error per pulse of $1/N = (0.186\pm0.008)\%$, corresponding to a fidelity of $0.9984$. 


In \FigRef{fig:RBM}(b) we plot the average error per gate versus pulse width $\tpw$, with the pulse period set to $T=3\,\tpw$. 
The predistorted pulses perform better for all pulse widths, showing that the pulse shapes have improved and again confirming the validity of our method for determining the transfer function $\mathcal{H}_\mathrm{int}$.  
The general trend is that the gate error is reduced for shorter pulses; this decreases the total time $t_\mathrm{total} = N\, T$ of the sequences, thereby reducing the errors due to loss of qubit coherence.  The relevant coherence time during the RBM sequence is a combination of $T_1$, $T_2$, and the coherence time during driven evolution; for simplicity we plot the expected error rate if the pulse errors were limited by $T_1=12 \us$ [dashed line in \FigRef{fig:RBM}(b)].  This limit is a factor of two lower than our best results, indicating that the predistorted pulses still contain some pulse imperfections.
 We speculate that parts of the remaining errors are due to in-phase pulse distortions, which are not targeted with the method presented here.  A similar scheme may be developed to investigate the in-phase errors independently.
Another complication is that, in our system, $T_1$ is strongly reduced when driving the qubit continuously at Rabi frequencies above $100\MHz$, probably due to local heating.  This may contribute to pulse errors for short pulses (where the drive amplitude $A \propto 1/\tpw$ becomes large).  At high drive amplitudes the Bloch-Siegert shift will also start to introduce deviations from the rotating-wave approximation in \EqRef{eq:Hr}.
%


To summarize, we have demonstrated a new technique of using a qubit to determine and correct microwave pulse imperfections, allowing us to generate single-qubit rotations with an average gate fidelity better than $99.8\%$. Even though there have been reports of superconducting qubits in 3D cavities with coherence times approaching $100\us$ \cite{Paik:2011,Rigetti:2012}, we note that we obtain a higher gate fidelity in our system because we are able to create shorter pulses. By encoding the pulse imperfections into a slow rotation when applying many pulses, we are able to detect distortions on a nanosecond timescale without the need of a fast detector.  This makes our method very general, and it can be applied to any quantum computing architecture where single-qubit gates are implemented by applying microwave pulses at the qubit frequency.


We thank X. Jin and P. Krantz for helpful discussions and K. Harrabi for assistance with device fabrication. This work was sponsored in part by the US Government, the Laboratory for Physical Sciences, the U.S. Army Research Office (W911NF-12-1-0036), the National Science Foundation (PHY-1005373), the Funding Program for World-Leading Innovative R\&D on Science and Technology (FIRST), NICT Commissioned Research, MEXT kakenhi 'Quantum Cybernetics', Project for Developing Innovation Systems of MEXT. Opinions, interpretations, conclusions and recommendations are those of the author(s) and are not necessarily endorsed by the US Government.

\bibliographystyle{apsrev4-1}
\bibliography{PulseCalibration}

\begin{thebibliography}{28}%
\makeatletter
\providecommand \@ifxundefined [1]{%
 \@ifx{#1\undefined}
}%
\providecommand \@ifnum [1]{%
 \ifnum #1\expandafter \@firstoftwo
 \else \expandafter \@secondoftwo
 \fi
}%
\providecommand \@ifx [1]{%
 \ifx #1\expandafter \@firstoftwo
 \else \expandafter \@secondoftwo
 \fi
}%
\providecommand \natexlab [1]{#1}%
\providecommand \enquote  [1]{``#1''}%
\providecommand \bibnamefont  [1]{#1}%
\providecommand \bibfnamefont [1]{#1}%
\providecommand \citenamefont [1]{#1}%
\providecommand \href@noop [0]{\@secondoftwo}%
\providecommand \href [0]{\begingroup \@sanitize@url \@href}%
\providecommand \@href[1]{\@@startlink{#1}\@@href}%
\providecommand \@@href[1]{\endgroup#1\@@endlink}%
\providecommand \@sanitize@url [0]{\catcode `\\12\catcode `\$12\catcode
  `\&12\catcode `\#12\catcode `\^12\catcode `\_12\catcode `\%12\relax}%
\providecommand \@@startlink[1]{}%
\providecommand \@@endlink[0]{}%
\providecommand \url  [0]{\begingroup\@sanitize@url \@url }%
\providecommand \@url [1]{\endgroup\@href {#1}{\urlprefix }}%
\providecommand \urlprefix  [0]{URL }%
\providecommand \Eprint [0]{\href }%
\providecommand \doibase [0]{http://dx.doi.org/}%
\providecommand \selectlanguage [0]{\@gobble}%
\providecommand \bibinfo  [0]{\@secondoftwo}%
\providecommand \bibfield  [0]{\@secondoftwo}%
\providecommand \translation [1]{[#1]}%
\providecommand \BibitemOpen [0]{}%
\providecommand \bibitemStop [0]{}%
\providecommand \bibitemNoStop [0]{.\EOS\space}%
\providecommand \EOS [0]{\spacefactor3000\relax}%
\providecommand \BibitemShut  [1]{\csname bibitem#1\endcsname}%
\let\auto@bib@innerbib\@empty
\bibitem [{\citenamefont {Bremner}\ \emph {et~al.}(2002)\citenamefont
  {Bremner}, \citenamefont {Dawson}, \citenamefont {Dodd}, \citenamefont
  {Gilchrist}, \citenamefont {Harrow}, \citenamefont {Mortimer}, \citenamefont
  {Nielsen},\ and\ \citenamefont {Osborne}}]{Bremner:2002}%
  \BibitemOpen
  \bibfield  {author} {\bibinfo {author} {\bibfnamefont {M.~J.}\ \bibnamefont
  {Bremner}}, \bibinfo {author} {\bibfnamefont {C.~M.}\ \bibnamefont {Dawson}},
  \bibinfo {author} {\bibfnamefont {J.~L.}\ \bibnamefont {Dodd}}, \bibinfo
  {author} {\bibfnamefont {A.}~\bibnamefont {Gilchrist}}, \bibinfo {author}
  {\bibfnamefont {A.~W.}\ \bibnamefont {Harrow}}, \bibinfo {author}
  {\bibfnamefont {D.}~\bibnamefont {Mortimer}}, \bibinfo {author}
  {\bibfnamefont {M.~A.}\ \bibnamefont {Nielsen}}, \ and\ \bibinfo {author}
  {\bibfnamefont {T.~J.}\ \bibnamefont {Osborne}},\ }\href@noop {} {\bibfield
  {journal} {\bibinfo  {journal} {Phys. Rev. Lett.}\ }\textbf {\bibinfo
  {volume} {89}},\ \bibinfo {pages} {247902} (\bibinfo {year}
  {2002})}\BibitemShut {NoStop}%
\bibitem [{\citenamefont {Lucero}\ \emph {et~al.}(2008)\citenamefont {Lucero},
  \citenamefont {Hofheinz}, \citenamefont {Ansmann}, \citenamefont {Bialczak},
  \citenamefont {Katz}, \citenamefont {Neeley}, \citenamefont {O'Connell},
  \citenamefont {Wang}, \citenamefont {Cleland},\ and\ \citenamefont
  {Martinis}}]{Lucero:2008}%
  \BibitemOpen
  \bibfield  {author} {\bibinfo {author} {\bibfnamefont {E.}~\bibnamefont
  {Lucero}}, \bibinfo {author} {\bibfnamefont {M.}~\bibnamefont {Hofheinz}},
  \bibinfo {author} {\bibfnamefont {M.}~\bibnamefont {Ansmann}}, \bibinfo
  {author} {\bibfnamefont {R.~C.}\ \bibnamefont {Bialczak}}, \bibinfo {author}
  {\bibfnamefont {N.}~\bibnamefont {Katz}}, \bibinfo {author} {\bibfnamefont
  {M.}~\bibnamefont {Neeley}}, \bibinfo {author} {\bibfnamefont {A.~D.}\
  \bibnamefont {O'Connell}}, \bibinfo {author} {\bibfnamefont {H.}~\bibnamefont
  {Wang}}, \bibinfo {author} {\bibfnamefont {A.~N.}\ \bibnamefont {Cleland}}, \
  and\ \bibinfo {author} {\bibfnamefont {J.~M.}\ \bibnamefont {Martinis}},\
  }\href@noop {} {\bibfield  {journal} {\bibinfo  {journal} {Phys. Rev. Lett.}\
  }\textbf {\bibinfo {volume} {100}},\ \bibinfo {pages} {247001} (\bibinfo
  {year} {2008})}\BibitemShut {NoStop}%
\bibitem [{\citenamefont {Chow}\ \emph {et~al.}(2009)\citenamefont {Chow},
  \citenamefont {Gambetta}, \citenamefont {Tornberg}, \citenamefont {Koch},
  \citenamefont {Bishop}, \citenamefont {Houck}, \citenamefont {Johnson},
  \citenamefont {Frunzio}, \citenamefont {Girvin},\ and\ \citenamefont
  {Schoelkopf}}]{Chow:2009}%
  \BibitemOpen
  \bibfield  {author} {\bibinfo {author} {\bibfnamefont {J.~M.}\ \bibnamefont
  {Chow}}, \bibinfo {author} {\bibfnamefont {J.~M.}\ \bibnamefont {Gambetta}},
  \bibinfo {author} {\bibfnamefont {L.}~\bibnamefont {Tornberg}}, \bibinfo
  {author} {\bibfnamefont {J.}~\bibnamefont {Koch}}, \bibinfo {author}
  {\bibfnamefont {L.~S.}\ \bibnamefont {Bishop}}, \bibinfo {author}
  {\bibfnamefont {A.~A.}\ \bibnamefont {Houck}}, \bibinfo {author}
  {\bibfnamefont {B.~R.}\ \bibnamefont {Johnson}}, \bibinfo {author}
  {\bibfnamefont {L.}~\bibnamefont {Frunzio}}, \bibinfo {author} {\bibfnamefont
  {S.~M.}\ \bibnamefont {Girvin}}, \ and\ \bibinfo {author} {\bibfnamefont
  {R.~J.}\ \bibnamefont {Schoelkopf}},\ }\href@noop {} {\bibfield  {journal}
  {\bibinfo  {journal} {Phys. Rev. Lett.}\ }\textbf {\bibinfo {volume} {102}},\
  \bibinfo {pages} {090502} (\bibinfo {year} {2009})}\BibitemShut {NoStop}%
\bibitem [{\citenamefont {Chow}\ \emph {et~al.}(2010)\citenamefont {Chow},
  \citenamefont {DiCarlo}, \citenamefont {Gambetta}, \citenamefont {Motzoi},
  \citenamefont {Frunzio}, \citenamefont {Girvin},\ and\ \citenamefont
  {Schoelkopf}}]{Chow:2010}%
  \BibitemOpen
  \bibfield  {author} {\bibinfo {author} {\bibfnamefont {J.~M.}\ \bibnamefont
  {Chow}}, \bibinfo {author} {\bibfnamefont {L.}~\bibnamefont {DiCarlo}},
  \bibinfo {author} {\bibfnamefont {J.~M.}\ \bibnamefont {Gambetta}}, \bibinfo
  {author} {\bibfnamefont {F.}~\bibnamefont {Motzoi}}, \bibinfo {author}
  {\bibfnamefont {L.}~\bibnamefont {Frunzio}}, \bibinfo {author} {\bibfnamefont
  {S.~M.}\ \bibnamefont {Girvin}}, \ and\ \bibinfo {author} {\bibfnamefont
  {R.~J.}\ \bibnamefont {Schoelkopf}},\ }\href@noop {} {\bibfield  {journal}
  {\bibinfo  {journal} {Phys. Rev. A}\ }\textbf {\bibinfo {volume} {82}},\
  \bibinfo {pages} {040305} (\bibinfo {year} {2010})}\BibitemShut {NoStop}%
\bibitem [{\citenamefont {Kerman}\ and\ \citenamefont
  {Oliver}(2008)}]{Kerman:2008}%
  \BibitemOpen
  \bibfield  {author} {\bibinfo {author} {\bibfnamefont {A.~J.}\ \bibnamefont
  {Kerman}}\ and\ \bibinfo {author} {\bibfnamefont {W.~D.}\ \bibnamefont
  {Oliver}},\ }\href@noop {} {\bibfield  {journal} {\bibinfo  {journal} {Phys.
  Rev. Lett.}\ }\textbf {\bibinfo {volume} {101}},\ \bibinfo {pages} {070501}
  (\bibinfo {year} {2008})}\BibitemShut {NoStop}%
\bibitem [{\citenamefont {Bialczak}\ \emph {et~al.}(2010)\citenamefont
  {Bialczak}, \citenamefont {Ansmann}, \citenamefont {Hofheinz}, \citenamefont
  {Lucero}, \citenamefont {Neeley}, \citenamefont {O'Connell}, \citenamefont
  {Sank}, \citenamefont {Wang}, \citenamefont {Wenner}, \citenamefont
  {Steffen}, \citenamefont {Cleland},\ and\ \citenamefont
  {Martinis}}]{Bialczak:2010}%
  \BibitemOpen
  \bibfield  {author} {\bibinfo {author} {\bibfnamefont {R.~C.}\ \bibnamefont
  {Bialczak}}, \bibinfo {author} {\bibfnamefont {M.}~\bibnamefont {Ansmann}},
  \bibinfo {author} {\bibfnamefont {M.}~\bibnamefont {Hofheinz}}, \bibinfo
  {author} {\bibfnamefont {E.}~\bibnamefont {Lucero}}, \bibinfo {author}
  {\bibfnamefont {M.}~\bibnamefont {Neeley}}, \bibinfo {author} {\bibfnamefont
  {A.~D.}\ \bibnamefont {O'Connell}}, \bibinfo {author} {\bibfnamefont
  {D.}~\bibnamefont {Sank}}, \bibinfo {author} {\bibfnamefont {H.}~\bibnamefont
  {Wang}}, \bibinfo {author} {\bibfnamefont {J.}~\bibnamefont {Wenner}},
  \bibinfo {author} {\bibfnamefont {M.}~\bibnamefont {Steffen}}, \bibinfo
  {author} {\bibfnamefont {A.~N.}\ \bibnamefont {Cleland}}, \ and\ \bibinfo
  {author} {\bibfnamefont {J.~M.}\ \bibnamefont {Martinis}},\ }\href@noop {}
  {\bibfield  {journal} {\bibinfo  {journal} {Nature Phys.}\ }\textbf {\bibinfo
  {volume} {6}},\ \bibinfo {pages} {409} (\bibinfo {year} {2010})}\BibitemShut
  {NoStop}%
\bibitem [{\citenamefont {Chow}\ \emph {et~al.}(2011)\citenamefont {Chow},
  \citenamefont {C{\'o}rcoles}, \citenamefont {Gambetta}, \citenamefont
  {Rigetti}, \citenamefont {Johnson}, \citenamefont {Smolin}, \citenamefont
  {Rozen}, \citenamefont {Keefe}, \citenamefont {Rothwell}, \citenamefont
  {Ketchen},\ and\ \citenamefont {Steffen}}]{Chow:2011}%
  \BibitemOpen
  \bibfield  {author} {\bibinfo {author} {\bibfnamefont {J.~M.}\ \bibnamefont
  {Chow}}, \bibinfo {author} {\bibfnamefont {A.~D.}\ \bibnamefont
  {C{\'o}rcoles}}, \bibinfo {author} {\bibfnamefont {J.~M.}\ \bibnamefont
  {Gambetta}}, \bibinfo {author} {\bibfnamefont {C.}~\bibnamefont {Rigetti}},
  \bibinfo {author} {\bibfnamefont {B.~R.}\ \bibnamefont {Johnson}}, \bibinfo
  {author} {\bibfnamefont {J.~A.}\ \bibnamefont {Smolin}}, \bibinfo {author}
  {\bibfnamefont {J.~R.}\ \bibnamefont {Rozen}}, \bibinfo {author}
  {\bibfnamefont {G.~A.}\ \bibnamefont {Keefe}}, \bibinfo {author}
  {\bibfnamefont {M.~B.}\ \bibnamefont {Rothwell}}, \bibinfo {author}
  {\bibfnamefont {M.~B.}\ \bibnamefont {Ketchen}}, \ and\ \bibinfo {author}
  {\bibfnamefont {M.}~\bibnamefont {Steffen}},\ }\href@noop {} {\bibfield
  {journal} {\bibinfo  {journal} {Phys. Rev. Lett.}\ }\textbf {\bibinfo
  {volume} {107}},\ \bibinfo {pages} {080502} (\bibinfo {year}
  {2011})}\BibitemShut {NoStop}%
\bibitem [{\citenamefont {Dewes}\ \emph {et~al.}(2012)\citenamefont {Dewes},
  \citenamefont {Ong}, \citenamefont {Schmitt}, \citenamefont {Lauro},
  \citenamefont {Boulant}, \citenamefont {Bertet}, \citenamefont {Vion},\ and\
  \citenamefont {Esteve}}]{Dewes:2012}%
  \BibitemOpen
  \bibfield  {author} {\bibinfo {author} {\bibfnamefont {A.}~\bibnamefont
  {Dewes}}, \bibinfo {author} {\bibfnamefont {F.~R.}\ \bibnamefont {Ong}},
  \bibinfo {author} {\bibfnamefont {V.}~\bibnamefont {Schmitt}}, \bibinfo
  {author} {\bibfnamefont {R.}~\bibnamefont {Lauro}}, \bibinfo {author}
  {\bibfnamefont {N.}~\bibnamefont {Boulant}}, \bibinfo {author} {\bibfnamefont
  {P.}~\bibnamefont {Bertet}}, \bibinfo {author} {\bibfnamefont
  {D.}~\bibnamefont {Vion}}, \ and\ \bibinfo {author} {\bibfnamefont
  {D.}~\bibnamefont {Esteve}},\ }\href@noop {} {\bibfield  {journal} {\bibinfo
  {journal} {Phys. Rev. Lett.}\ }\textbf {\bibinfo {volume} {108}},\ \bibinfo
  {pages} {057002} (\bibinfo {year} {2012})}\BibitemShut {NoStop}%
\bibitem [{\citenamefont {Gustavsson}\ \emph {et~al.}(2012)\citenamefont
  {Gustavsson}, \citenamefont {Yan}, \citenamefont {Bylander}, \citenamefont
  {Yoshihara}, \citenamefont {Nakamura}, \citenamefont {Orlando},\ and\
  \citenamefont {Oliver}}]{GustavssonTLS:2012}%
  \BibitemOpen
  \bibfield  {author} {\bibinfo {author} {\bibfnamefont {S.}~\bibnamefont
  {Gustavsson}}, \bibinfo {author} {\bibfnamefont {F.}~\bibnamefont {Yan}},
  \bibinfo {author} {\bibfnamefont {J.}~\bibnamefont {Bylander}}, \bibinfo
  {author} {\bibfnamefont {F.}~\bibnamefont {Yoshihara}}, \bibinfo {author}
  {\bibfnamefont {Y.}~\bibnamefont {Nakamura}}, \bibinfo {author}
  {\bibfnamefont {T.~P.}\ \bibnamefont {Orlando}}, \ and\ \bibinfo {author}
  {\bibfnamefont {W.~D.}\ \bibnamefont {Oliver}},\ }\href@noop {} {\bibfield
  {journal} {\bibinfo  {journal} {Phys. Rev. Lett.}\ }\textbf {\bibinfo
  {volume} {109}},\ \bibinfo {pages} {010502} (\bibinfo {year}
  {2012})}\BibitemShut {NoStop}%
\bibitem [{\citenamefont {Steffen}\ \emph {et~al.}(2012)\citenamefont
  {Steffen}, \citenamefont {Silva}, \citenamefont {Fedorov}, \citenamefont
  {Baur},\ and\ \citenamefont {Wallraff}}]{Steffen:2012}%
  \BibitemOpen
  \bibfield  {author} {\bibinfo {author} {\bibfnamefont {L.}~\bibnamefont
  {Steffen}}, \bibinfo {author} {\bibfnamefont {M.~D.}\ \bibnamefont {Silva}},
  \bibinfo {author} {\bibfnamefont {A.}~\bibnamefont {Fedorov}}, \bibinfo
  {author} {\bibfnamefont {M.}~\bibnamefont {Baur}}, \ and\ \bibinfo {author}
  {\bibfnamefont {A.}~\bibnamefont {Wallraff}},\ }\href@noop {} {\bibfield
  {journal} {\bibinfo  {journal} {Phys. Rev. Lett.}\ }\textbf {\bibinfo
  {volume} {108}},\ \bibinfo {pages} {260506} (\bibinfo {year}
  {2012})}\BibitemShut {NoStop}%
\bibitem [{\citenamefont {Chow}\ \emph {et~al.}(2012)\citenamefont {Chow},
  \citenamefont {Gambetta}, \citenamefont {C{\'o}rcoles}, \citenamefont
  {Merkel}, \citenamefont {Smolin}, \citenamefont {Rigetti}, \citenamefont
  {Poletto}, \citenamefont {Keefe}, \citenamefont {Rothwell}, \citenamefont
  {Rozen}, \citenamefont {Ketchen},\ and\ \citenamefont {Steffen}}]{Chow:2012}%
  \BibitemOpen
  \bibfield  {author} {\bibinfo {author} {\bibfnamefont {J.~M.}\ \bibnamefont
  {Chow}}, \bibinfo {author} {\bibfnamefont {J.~M.}\ \bibnamefont {Gambetta}},
  \bibinfo {author} {\bibfnamefont {A.~D.}\ \bibnamefont {C{\'o}rcoles}},
  \bibinfo {author} {\bibfnamefont {S.~T.}\ \bibnamefont {Merkel}}, \bibinfo
  {author} {\bibfnamefont {J.~A.}\ \bibnamefont {Smolin}}, \bibinfo {author}
  {\bibfnamefont {C.}~\bibnamefont {Rigetti}}, \bibinfo {author} {\bibfnamefont
  {S.}~\bibnamefont {Poletto}}, \bibinfo {author} {\bibfnamefont {G.~A.}\
  \bibnamefont {Keefe}}, \bibinfo {author} {\bibfnamefont {M.~B.}\ \bibnamefont
  {Rothwell}}, \bibinfo {author} {\bibfnamefont {J.~R.}\ \bibnamefont {Rozen}},
  \bibinfo {author} {\bibfnamefont {M.~B.}\ \bibnamefont {Ketchen}}, \ and\
  \bibinfo {author} {\bibfnamefont {M.}~\bibnamefont {Steffen}},\ }\href@noop
  {} {\bibfield  {journal} {\bibinfo  {journal} {Phys. Rev. Lett.}\ }\textbf
  {\bibinfo {volume} {109}},\ \bibinfo {pages} {060501} (\bibinfo {year}
  {2012})}\BibitemShut {NoStop}%
\bibitem [{\citenamefont {Knill}(2005)}]{Knill:2005}%
  \BibitemOpen
  \bibfield  {author} {\bibinfo {author} {\bibfnamefont {E.}~\bibnamefont
  {Knill}},\ }\href@noop {} {\bibfield  {journal} {\bibinfo  {journal}
  {Nature}\ }\textbf {\bibinfo {volume} {434}},\ \bibinfo {pages} {39}
  (\bibinfo {year} {2005})}\BibitemShut {NoStop}%
\bibitem [{\citenamefont {DiVincenzo}(2009)}]{DiVincenzo:2009}%
  \BibitemOpen
  \bibfield  {author} {\bibinfo {author} {\bibfnamefont {D.~P.}\ \bibnamefont
  {DiVincenzo}},\ }\href@noop {} {\bibfield  {journal} {\bibinfo  {journal}
  {Phys. Scr.}\ }\textbf {\bibinfo {volume} {T137}},\ \bibinfo {pages} {014020}
  (\bibinfo {year} {2009})}\BibitemShut {NoStop}%
\bibitem [{\citenamefont {Bylander}\ \emph {et~al.}(2009)\citenamefont
  {Bylander}, \citenamefont {Rudner}, \citenamefont {Shytov}, \citenamefont
  {Valenzuela}, \citenamefont {Berns}, \citenamefont {Berggren}, \citenamefont
  {Levitov},\ and\ \citenamefont {Oliver}}]{Bylander:2009}%
  \BibitemOpen
  \bibfield  {author} {\bibinfo {author} {\bibfnamefont {J.}~\bibnamefont
  {Bylander}}, \bibinfo {author} {\bibfnamefont {M.~S.}\ \bibnamefont
  {Rudner}}, \bibinfo {author} {\bibfnamefont {A.~V.}\ \bibnamefont {Shytov}},
  \bibinfo {author} {\bibfnamefont {S.~O.}\ \bibnamefont {Valenzuela}},
  \bibinfo {author} {\bibfnamefont {D.~M.}\ \bibnamefont {Berns}}, \bibinfo
  {author} {\bibfnamefont {K.~K.}\ \bibnamefont {Berggren}}, \bibinfo {author}
  {\bibfnamefont {L.~S.}\ \bibnamefont {Levitov}}, \ and\ \bibinfo {author}
  {\bibfnamefont {W.~D.}\ \bibnamefont {Oliver}},\ }\href@noop {} {\bibfield
  {journal} {\bibinfo  {journal} {Phys. Rev. B}\ }\textbf {\bibinfo {volume}
  {80}},\ \bibinfo {pages} {220506} (\bibinfo {year} {2009})}\BibitemShut
  {NoStop}%
\bibitem [{\citenamefont {Knill}\ \emph {et~al.}(2008)\citenamefont {Knill},
  \citenamefont {Leibfried}, \citenamefont {Reichle}, \citenamefont {Britton},
  \citenamefont {Blakestad}, \citenamefont {Jost}, \citenamefont {Langer},
  \citenamefont {Ozeri}, \citenamefont {Seidelin},\ and\ \citenamefont
  {Wineland}}]{Knill:2008}%
  \BibitemOpen
  \bibfield  {author} {\bibinfo {author} {\bibfnamefont {E.}~\bibnamefont
  {Knill}}, \bibinfo {author} {\bibfnamefont {D.}~\bibnamefont {Leibfried}},
  \bibinfo {author} {\bibfnamefont {R.}~\bibnamefont {Reichle}}, \bibinfo
  {author} {\bibfnamefont {J.}~\bibnamefont {Britton}}, \bibinfo {author}
  {\bibfnamefont {R.~B.}\ \bibnamefont {Blakestad}}, \bibinfo {author}
  {\bibfnamefont {J.~D.}\ \bibnamefont {Jost}}, \bibinfo {author}
  {\bibfnamefont {C.}~\bibnamefont {Langer}}, \bibinfo {author} {\bibfnamefont
  {R.}~\bibnamefont {Ozeri}}, \bibinfo {author} {\bibfnamefont
  {S.}~\bibnamefont {Seidelin}}, \ and\ \bibinfo {author} {\bibfnamefont
  {D.~J.}\ \bibnamefont {Wineland}},\ }\href@noop {} {\bibfield  {journal}
  {\bibinfo  {journal} {Phys. Rev. A}\ }\textbf {\bibinfo {volume} {77}},\
  \bibinfo {pages} {012307} (\bibinfo {year} {2008})}\BibitemShut {NoStop}%
\bibitem [{\citenamefont {Mooij}\ \emph {et~al.}(1999)\citenamefont {Mooij},
  \citenamefont {Orlando}, \citenamefont {Levitov}, \citenamefont {Tian},
  \citenamefont {van~der Wal},\ and\ \citenamefont {Lloyd}}]{Mooij:1999}%
  \BibitemOpen
  \bibfield  {author} {\bibinfo {author} {\bibfnamefont {J.}~\bibnamefont
  {Mooij}}, \bibinfo {author} {\bibfnamefont {T.}~\bibnamefont {Orlando}},
  \bibinfo {author} {\bibfnamefont {L.}~\bibnamefont {Levitov}}, \bibinfo
  {author} {\bibfnamefont {L.}~\bibnamefont {Tian}}, \bibinfo {author}
  {\bibfnamefont {C.~H.}\ \bibnamefont {van~der Wal}}, \ and\ \bibinfo {author}
  {\bibfnamefont {S.}~\bibnamefont {Lloyd}},\ }\href@noop {} {\bibfield
  {journal} {\bibinfo  {journal} {Science}\ }\textbf {\bibinfo {volume}
  {285}},\ \bibinfo {pages} {1036} (\bibinfo {year} {1999})}\BibitemShut
  {NoStop}%
\bibitem [{\citenamefont {Bylander}\ \emph {et~al.}(2011)\citenamefont
  {Bylander}, \citenamefont {Gustavsson}, \citenamefont {Yan}, \citenamefont
  {Yoshihara}, \citenamefont {Harrabi}, \citenamefont {Fitch}, \citenamefont
  {Cory}, \citenamefont {Nakamura}, \citenamefont {Tsai},\ and\ \citenamefont
  {Oliver}}]{Bylander:2011}%
  \BibitemOpen
  \bibfield  {author} {\bibinfo {author} {\bibfnamefont {J.}~\bibnamefont
  {Bylander}}, \bibinfo {author} {\bibfnamefont {S.}~\bibnamefont
  {Gustavsson}}, \bibinfo {author} {\bibfnamefont {F.}~\bibnamefont {Yan}},
  \bibinfo {author} {\bibfnamefont {F.}~\bibnamefont {Yoshihara}}, \bibinfo
  {author} {\bibfnamefont {K.}~\bibnamefont {Harrabi}}, \bibinfo {author}
  {\bibfnamefont {G.}~\bibnamefont {Fitch}}, \bibinfo {author} {\bibfnamefont
  {D.~G.}\ \bibnamefont {Cory}}, \bibinfo {author} {\bibfnamefont
  {Y.}~\bibnamefont {Nakamura}}, \bibinfo {author} {\bibfnamefont {J.~S.}\
  \bibnamefont {Tsai}}, \ and\ \bibinfo {author} {\bibfnamefont {W.~D.}\
  \bibnamefont {Oliver}},\ }\href@noop {} {\bibfield  {journal} {\bibinfo
  {journal} {Nature Phys.}\ }\textbf {\bibinfo {volume} {7}},\ \bibinfo {pages}
  {565} (\bibinfo {year} {2011})}\BibitemShut {NoStop}%
\bibitem [{\citenamefont {Chiorescu}\ \emph {et~al.}(2003)\citenamefont
  {Chiorescu}, \citenamefont {Nakamura}, \citenamefont {Harmans},\ and\
  \citenamefont {Mooij}}]{Chiorescu:2003}%
  \BibitemOpen
  \bibfield  {author} {\bibinfo {author} {\bibfnamefont {I.}~\bibnamefont
  {Chiorescu}}, \bibinfo {author} {\bibfnamefont {Y.}~\bibnamefont {Nakamura}},
  \bibinfo {author} {\bibfnamefont {C.~J. P.~M.}\ \bibnamefont {Harmans}}, \
  and\ \bibinfo {author} {\bibfnamefont {J.~E.}\ \bibnamefont {Mooij}},\
  }\href@noop {} {\bibfield  {journal} {\bibinfo  {journal} {Science}\ }\textbf
  {\bibinfo {volume} {299}},\ \bibinfo {pages} {1869} (\bibinfo {year}
  {2003})}\BibitemShut {NoStop}%
\bibitem [{\citenamefont {Hofheinz}\ \emph {et~al.}(2009)\citenamefont
  {Hofheinz}, \citenamefont {Wang}, \citenamefont {Ansmann}, \citenamefont
  {Bialczak}, \citenamefont {Lucero}, \citenamefont {Neeley}, \citenamefont
  {O'Connell}, \citenamefont {Sank}, \citenamefont {Wenner}, \citenamefont
  {Martinis},\ and\ \citenamefont {Cleland}}]{Hofheinz:2009}%
  \BibitemOpen
  \bibfield  {author} {\bibinfo {author} {\bibfnamefont {M.}~\bibnamefont
  {Hofheinz}}, \bibinfo {author} {\bibfnamefont {H.}~\bibnamefont {Wang}},
  \bibinfo {author} {\bibfnamefont {M.}~\bibnamefont {Ansmann}}, \bibinfo
  {author} {\bibfnamefont {R.~C.}\ \bibnamefont {Bialczak}}, \bibinfo {author}
  {\bibfnamefont {E.}~\bibnamefont {Lucero}}, \bibinfo {author} {\bibfnamefont
  {M.}~\bibnamefont {Neeley}}, \bibinfo {author} {\bibfnamefont {A.~D.}\
  \bibnamefont {O'Connell}}, \bibinfo {author} {\bibfnamefont {D.}~\bibnamefont
  {Sank}}, \bibinfo {author} {\bibfnamefont {J.}~\bibnamefont {Wenner}},
  \bibinfo {author} {\bibfnamefont {J.~M.}\ \bibnamefont {Martinis}}, \ and\
  \bibinfo {author} {\bibfnamefont {A.~N.}\ \bibnamefont {Cleland}},\
  }\href@noop {} {\bibfield  {journal} {\bibinfo  {journal} {Nature}\ }\textbf
  {\bibinfo {volume} {459}},\ \bibinfo {pages} {546} (\bibinfo {year}
  {2009})}\BibitemShut {NoStop}%
\bibitem [{\citenamefont {Johnson}(2011)}]{JohnsonThesis:2011}%
  \BibitemOpen
  \bibfield  {author} {\bibinfo {author} {\bibfnamefont {B.~R.}\ \bibnamefont
  {Johnson}},\ }\href@noop {} {\bibfield  {journal} {\bibinfo  {journal} {PhD
  Thesis, Yale University}\ } (\bibinfo {year} {2011})}\BibitemShut {NoStop}%
\bibitem [{Note1()}]{Note1}%
  \BibitemOpen
  \bibinfo {note} {See online supplementary material S1.}\BibitemShut {Stop}%
\bibitem [{\citenamefont {Steffen}\ \emph {et~al.}(2006)\citenamefont
  {Steffen}, \citenamefont {Ansmann}, \citenamefont {McDermott}, \citenamefont
  {Katz}, \citenamefont {Bialczak}, \citenamefont {Lucero}, \citenamefont
  {Neeley}, \citenamefont {Weig}, \citenamefont {Cleland},\ and\ \citenamefont
  {Martinis}}]{Steffen:2006}%
  \BibitemOpen
  \bibfield  {author} {\bibinfo {author} {\bibfnamefont {M.}~\bibnamefont
  {Steffen}}, \bibinfo {author} {\bibfnamefont {M.}~\bibnamefont {Ansmann}},
  \bibinfo {author} {\bibfnamefont {R.}~\bibnamefont {McDermott}}, \bibinfo
  {author} {\bibfnamefont {N.}~\bibnamefont {Katz}}, \bibinfo {author}
  {\bibfnamefont {R.~C.}\ \bibnamefont {Bialczak}}, \bibinfo {author}
  {\bibfnamefont {E.}~\bibnamefont {Lucero}}, \bibinfo {author} {\bibfnamefont
  {M.}~\bibnamefont {Neeley}}, \bibinfo {author} {\bibfnamefont {E.~M.}\
  \bibnamefont {Weig}}, \bibinfo {author} {\bibfnamefont {A.~N.}\ \bibnamefont
  {Cleland}}, \ and\ \bibinfo {author} {\bibfnamefont {J.~M.}\ \bibnamefont
  {Martinis}},\ }\href@noop {} {\bibfield  {journal} {\bibinfo  {journal}
  {Phys. Rev. Lett.}\ }\textbf {\bibinfo {volume} {97}},\ \bibinfo {pages}
  {050502} (\bibinfo {year} {2006})}\BibitemShut {NoStop}%
\bibitem [{\citenamefont {Hahn}(1950)}]{Hahn:1950}%
  \BibitemOpen
  \bibfield  {author} {\bibinfo {author} {\bibfnamefont {E.}~\bibnamefont
  {Hahn}},\ }\href@noop {} {\bibfield  {journal} {\bibinfo  {journal} {Phys.
  Rev.}\ }\textbf {\bibinfo {volume} {80}},\ \bibinfo {pages} {580} (\bibinfo
  {year} {1950})}\BibitemShut {NoStop}%
\bibitem [{Note2()}]{Note2}%
  \BibitemOpen
  \bibinfo {note} {See online supplementary material S2.}\BibitemShut {Stop}%
\bibitem [{Note3()}]{Note3}%
  \BibitemOpen
  \bibinfo {note} {See online supplementary material S3.}\BibitemShut {Stop}%
\bibitem [{Note4()}]{Note4}%
  \BibitemOpen
  \bibinfo {note} {See online supplementary material S4.}\BibitemShut {Stop}%
\bibitem [{\citenamefont {Paik}\ \emph {et~al.}(2011)\citenamefont {Paik},
  \citenamefont {Schuster}, \citenamefont {Bishop}, \citenamefont {Kirchmair},
  \citenamefont {Catelani}, \citenamefont {Sears}, \citenamefont {Johnson},
  \citenamefont {Reagor}, \citenamefont {Frunzio}, \citenamefont {Glazman},
  \citenamefont {Girvin}, \citenamefont {Devoret},\ and\ \citenamefont
  {Schoelkopf}}]{Paik:2011}%
  \BibitemOpen
  \bibfield  {author} {\bibinfo {author} {\bibfnamefont {H.}~\bibnamefont
  {Paik}}, \bibinfo {author} {\bibfnamefont {D.}~\bibnamefont {Schuster}},
  \bibinfo {author} {\bibfnamefont {L.}~\bibnamefont {Bishop}}, \bibinfo
  {author} {\bibfnamefont {G.}~\bibnamefont {Kirchmair}}, \bibinfo {author}
  {\bibfnamefont {G.}~\bibnamefont {Catelani}}, \bibinfo {author}
  {\bibfnamefont {A.}~\bibnamefont {Sears}}, \bibinfo {author} {\bibfnamefont
  {B.}~\bibnamefont {Johnson}}, \bibinfo {author} {\bibfnamefont
  {M.}~\bibnamefont {Reagor}}, \bibinfo {author} {\bibfnamefont
  {L.}~\bibnamefont {Frunzio}}, \bibinfo {author} {\bibfnamefont
  {L.}~\bibnamefont {Glazman}}, \bibinfo {author} {\bibfnamefont
  {S.}~\bibnamefont {Girvin}}, \bibinfo {author} {\bibfnamefont
  {M.}~\bibnamefont {Devoret}}, \ and\ \bibinfo {author} {\bibfnamefont
  {R.}~\bibnamefont {Schoelkopf}},\ }\href@noop {} {\bibfield  {journal}
  {\bibinfo  {journal} {Phys. Rev. Lett.}\ }\textbf {\bibinfo {volume} {107}},\
  \bibinfo {pages} {240501} (\bibinfo {year} {2011})}\BibitemShut {NoStop}%
\bibitem [{\citenamefont {Rigetti}\ \emph {et~al.}(2012)\citenamefont
  {Rigetti}, \citenamefont {Gambetta}, \citenamefont {Poletto}, \citenamefont
  {Plourde}, \citenamefont {Chow}, \citenamefont {C{\'o}rcoles}, \citenamefont
  {Smolin}, \citenamefont {Merkel}, \citenamefont {Rozen}, \citenamefont
  {Keefe}, \citenamefont {Rothwell}, \citenamefont {Ketchen},\ and\
  \citenamefont {Steffen}}]{Rigetti:2012}%
  \BibitemOpen
  \bibfield  {author} {\bibinfo {author} {\bibfnamefont {C.}~\bibnamefont
  {Rigetti}}, \bibinfo {author} {\bibfnamefont {J.~M.}\ \bibnamefont
  {Gambetta}}, \bibinfo {author} {\bibfnamefont {S.}~\bibnamefont {Poletto}},
  \bibinfo {author} {\bibfnamefont {B.~L.~T.}\ \bibnamefont {Plourde}},
  \bibinfo {author} {\bibfnamefont {J.~M.}\ \bibnamefont {Chow}}, \bibinfo
  {author} {\bibfnamefont {A.~D.}\ \bibnamefont {C{\'o}rcoles}}, \bibinfo
  {author} {\bibfnamefont {J.~A.}\ \bibnamefont {Smolin}}, \bibinfo {author}
  {\bibfnamefont {S.~T.}\ \bibnamefont {Merkel}}, \bibinfo {author}
  {\bibfnamefont {J.~R.}\ \bibnamefont {Rozen}}, \bibinfo {author}
  {\bibfnamefont {G.~A.}\ \bibnamefont {Keefe}}, \bibinfo {author}
  {\bibfnamefont {M.~B.}\ \bibnamefont {Rothwell}}, \bibinfo {author}
  {\bibfnamefont {M.~B.}\ \bibnamefont {Ketchen}}, \ and\ \bibinfo {author}
  {\bibfnamefont {M.}~\bibnamefont {Steffen}},\ }\href@noop {} {\bibfield
  {journal} {\bibinfo  {journal} {Phys. Rev. B}\ }\textbf {\bibinfo {volume}
  {86}},\ \bibinfo {pages} {100506} (\bibinfo {year} {2012})}\BibitemShut
  {NoStop}%
\end{thebibliography}%

\clearpage

\begin{widetext} 

\section*{Supplement S1: \\Correcting pulse distortions due to instrument imperfections} \label{sec:s1}
The microwave pulses that implement qubit rotations are generated by creating in-phase (I) and quadrature (Q) pulse envelopes using a Tektronix 5014 arbitrary waveform generator (AWG), and sending them to the internal IQ mixer of an Agilent 8267D microwave generator.  The AWG has a sampling rate of $1.2\GSs$ and an analog bandwidth of $300\MHz$.  To ensure that the pulses we send to the sample are initially free from distortions, we use a high-speed oscilloscope (Tektronix DPO71254B, with maximal sampling rate $40\GSs$ and an analog bandwidth of $12.5\GHz$) to sample the microwave pulses and extract the pulse envelopes.  This enables us to infer the transfer function of the instrument, from which we use signal processing techniques to correct for imperfections in the AWG and in the IQ mixers of the microwave generator. 

In general, a transfer function $h(t)$ takes an input signal $x(t)$ and converts it into the output signal $y(t)$.  In the time domain, the transfer function has the form of a convolution $y(t) = h(t)*x(t)$; by going to Fourier space we get
\begin{equation}
Y(f) = \mathcal{H}(f) \,X(f).
\end{equation}
or 
\begin{equation} \label{eq:H}
\mathcal{H}(f) =Y(f)/X(f). 
\end{equation}
We can thus determine $\Ht(f)$ by measuring the response $Y(f)$ for a known input signal $X(f)$.  If $x(t)$ is a delta function, $X(f) = 1$, giving $\Ht(f) = Y(f)$, and the transfer function is equal to the impulse response of the system.
For technical reasons, we found it easier to measure the step response instead of the impulse response, but the impulse response can easily be determined by taking the time derivative of the measured step response.
To find the combined transfer function of the AWG and the IQ mixers, we used the following procedure (similar to Ref.~\cite{JohnsonThesis:2011}):

\begin{enumerate}
\item Generate a signal at the qubit frequency ($\fmw=5.4\GHz$) with the microwave generator and send it to the mixer input.  
\item Apply a $5\us$ long square pulse to the mixer I channel, apply zeros to the mixer Q channel.  The relatively long pulse duration is necessary to make sure the transfer functions captures the slow transients in the AWG output voltage.
\item Sample the signal $g(t)$ coming out of the mixer at $40\GSs$ using the high-speed oscilloscope.
\item Determine the in-phase [$\Ai(t)$] and quadrature [$\Aq(t)$] components of the sampled signal. We do this by multiplying the sampled signal with a cosine (for $\Ai$) or sine (for $\Aq$) and integrate over one period $T = 1/\fmw$ to get the pulse envelopes. We have
\begin{eqnarray}
\Ai(t) & = & \frac{2}{T} \int_t^{t+T} g(t') \cos\left( 2\pi \fmw t' \right) \, dt' \\
\Aq(t) & = & -\frac{2}{T} \int_t^{t+T} g(t') \sin\left( 2\pi \fmw t' \right) \, dt'.
\end{eqnarray}
\item Resample the envelopes $\Ai(t)$ and $\Aq(t)$ at the sample rate of the AWG (in our case $1.2\GSs$).
\item Convert the step responses $\Ai(t)$ and $\Aq(t)$ to impulse responses by taking the numerical derivative.
\item Create the complex impulse response function $y(t) = (d/dt) \left( \Ai(t) + i \,\Aq(t) \right)$.
\item Calculate the Fourier transform of $y(t)$ to get the transfer function $\Ht(f) = Y(f)$. Note that since $y(t)$ is complex, $\Ht(f)$ does not generally have a symmetric spectrum.
\item Create the inverse $\Hi(f) = 1/\Ht(f)$.
\end{enumerate}
Once the transfer function has been determined, we use the inverse $\Hi$ to predistort the input signal.  The system will then generate the desired output signal $y(t) = h(t) * h^{-1}(t) * x(t) = x(t)$. Practically, we implement the following procedure:
\begin{enumerate}
\item Generate (in MATLAB) the in-phase [$\Ai(t)$] and quadrature [$\Aq(t)$] envelopes of the desired pulse sequence, sampled at the sample rate of the AWG.
\item Construct the complex input signal $x(t) = \Ai(t) + i \,\Aq(t)$.
\item Take the discrete-time Fourier transform of $x(t)$ to get $X(f)$.
\item Predistort the signal by multiplying $\Hi(f)$ with $X(f)$.
\item Return to the time domain by taking the inverse Fourier transform of $\Hi(f)X(f)$.
\item The real part of the resulting signal is the predistorted in-phase component, the imaginary part is the quadrature component.
\item Generate the two signals in the AWG and send them to the I and Q ports of the mixer.
\end{enumerate}
\begin{figure}[b!]
\centering
\includegraphics[width=\linewidth]{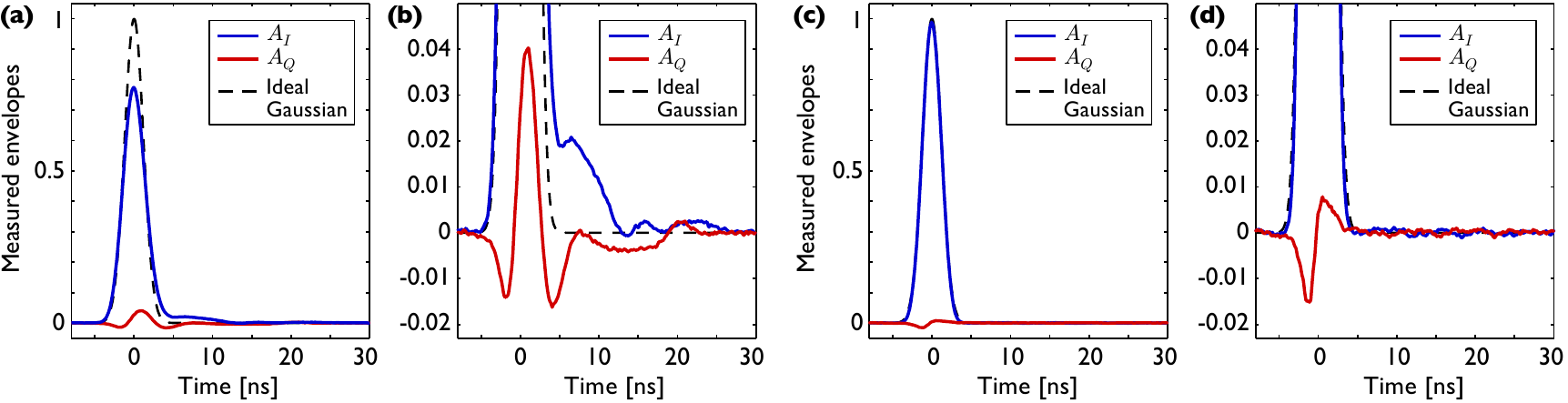}
\caption{(a) Measured in-phase (I) and quadrature (Q) components of a $3\ns$ long Gaussian microwave pulse. The black dashed line shows the ideal Gaussian. (b) Magnification of the low-amplitude region of (a).  (c-d) Same as panels (a-b), but here the pulse was generated using the predistortion algorithm.  The pulse distortion that is strongly visible in (b) is essentially gone in panel (d).
}
\label{fig:Pulses}
\end{figure}
Figure~\ref{fig:Pulses} demonstrates the improvement of the pulse shape after implementing the predistortion algorithm. In Fig.~\ref{fig:Pulses}(a), we plot the extracted pulse envelopes of an intended $3\ns$ wide Gaussian pulse, created without any predistortion,  while Fig.~\ref{fig:Pulses}(c) shows the same pulse generated using predistortion.  Figures~\ref{fig:Pulses}(b) and (d) are magnifications of low-amplitude regions of Figs~\ref{fig:Pulses}(a) and (c), respectively. For the predistorted version, the ringing after the pulse and the quadrature components visible in \FigRef{fig:Pulses}(b) are practically gone.  Also note that the pulse without predistortion in \FigRef{fig:Pulses}(a) does not reach up to the intended unity amplitude, which the predistorted curve does.

\section*{Supplement S2: \\Defining the matrix $\mathbf{M}$} \label{sec:s2}
From Eq.~(2) in the main text, we have
\begin{equation}
\theta_m = \Delta t \left[\sum_{n=1}^m Q_n - \sum_{m+1}^{2m} Q_n +\sum_{2m+1}^{3m} Q_n -  \sum_{3m+1}^{4m} Q_n +\cdots \right].
\label{eq:R1}  
\end{equation}
To see how this can be written in matrix form, we consider a specific example where the vectors $\vec{Q}$ and $\vec{\theta}$  each have $N=10$ elements. For $m=1$ (where the period between $\pi$ pulses is one element), the total rotation $\theta_1$ becomes
\begin{equation}
\theta_1 = \Delta t \left[ Q_1 -Q_2 + Q_3 - Q_4 +Q_5 - Q_6 + Q_7 - Q_8 + Q_9 - Q_{10} \right].
\label{eq:R1_1}  
\end{equation}
and for $m=2$ we get
\begin{equation}
\theta_2 = \Delta t \left[ Q_1 +Q_2 - Q_3 - Q_4 +Q_5 + Q_6 - Q_7 - Q_8 + Q_9 + Q_{10} \right].
\label{eq:R1_2}  
\end{equation}
Written in this form, the matrix nature of \EqRef{eq:R1} becomes more apparent.  Writing out the full matrix, we have
\begin{equation}
\vec{\theta} = \Delta t \left[\begin{array}{cccccccccc}+1 & -1 & +1 & -1 & +1 & -1 & +1 & -1 & +1 & -1 \\+1 & +1 & -1 & -1 & +1 & +1 & -1 & -1 & +1 & +1 \\+1 & +1 & +1 & -1 & -1 & -1 & +1 & +1 & +1 & -1 \\+1 & +1 & +1 & +1 & -1 & -1 & -1 & -1 & +1 & +1 \\+1 & +1 & +1 & +1 & +1 & -1 & -1 & -1 & -1 & -1 \\+1 & +1 & +1 & +1 & +1 & +1 & -1 & -1 & -1 & -1 \\+1 & +1 & +1 & +1 & +1 & +1 & +1 & -1 & -1 & -1 \\+1 & +1 & +1 & +1 & +1 & +1 & +1 & +1 & -1 & -1 \\+1 & +1 & +1 & +1 & +1 & +1 & +1 & +1 & +1 & -1 \\+1 & +1 & +1 & +1 & +1 & +1 & +1 & +1 & +1 & +1\end{array}\right]\cdot \vec{Q}
\label{eq:M1}  
\end{equation}

\section*{Supplement S3: \\Modifying the matrix $\mathbf{M}$ to incorporate finite pulse widths} \label{sec:s2}
Equation~(2) in the main text and the matrix $\mathbf{M}$ in \EqRef{eq:M1} above both assume that the $\pi$ pulses reversing the rotation directions are instantaneous.  This is not the case in the experiment, where we use Gaussian pulses with $\tpw = 2.5\ns$.  During the pulses, the qubit is strongly driven around the $x$-axis, and we can neglect the $y$-rotations from $\Aq$ at the $\tpw/\Delta t = 3$ discrete time steps when $\Ai\gg \Aq$.  This means that we can not extract any information about $Q$ during the pulse, and the vectors will be limited to $\vec{Q} = [Q_4,Q_5,\cdots, Q_{N}]$ and $\vec{\theta} = [\theta_4,\theta_5,\cdots, \theta_{N}]$.  
We need to rewrite Eq.~(2) in the main text as
\begin{equation}
\theta_m = \Delta t \left[\sum_{n=4}^m Q_n - \sum_{m+4}^{2m} Q_n +\sum_{2m+4}^{3m} Q_n - \sum_{3m+4}^{4m} Q_n  +\cdots \right].
\label{eq:R2}  
\end{equation} 
The changes are most easily visualized by writing \EqRef{eq:R2} in matrix form (again for the case $N=10$):
\begin{equation}
\vec{\theta} = \Delta t \left[\begin{array}{cccccccccc}  &  &  &  &  &  &  &  &  &  \\ &  &  &  &  &  &  &  &  &  \\ &  &  & &  & &  &  &  &  \\ &  &  & +1 & 0 & 0 & 0 & -1 & 0 & 0 \\ &  &  & +1 & +1 & 0 & 0 & 0 & -1 & -1 \\ &  &  & +1 & +1 & +1 & 0 & 0 & 0 & -1 \\ &  &  & +1 & +1 & +1 & +1 & 0 & 0 & 0 \\ &  &  & +1 & +1 & +1 & +1 & +1 & 0 & 0 \\ &  &  & +1 & +1 & +1 & +1 & +1 & +1 & 0 \\ &  &  & +1 & +1 & +1 & +1 & +1 & +1 & +1\end{array}\right]\cdot \vec{Q}
\label{eq:M2}  
\end{equation}
Here, we have removed the entries for $n,m=1,2,3$ (since the vectors are three elements shorter).  Also, we have include zeros at the positions where the $\pi$ pulses occur in $\Ai$ (since during the $\pi$ pulses, $\Ai\gg \Aq$, and the quadrature component is very inefficient in driving rotations around $y$).

\section*{Supplement S4: \\Predistorting using the response extracted from the qubit quadrature rotations}
The procedure for predistorting signals based on the qubit response is very similar to the sequence described in section S1, with some slight differences in how we determine the transfer function $\Hi$.  After having calculated the quadrature component $\vec{Q}$ as described in the main text, we proceed as follows:
\begin{enumerate}
\item Construct the output signal $y(t) = \vec{I}+ i\, \vec{Q}$ from the two vectors $\vec{I}$ and $\vec{Q}$. Here, $\vec{Q}$ is the extracted quadrature component, while we set $\vec{I}$ to be an ideal Gaussian with pulse width $\tpw=2.5\ns$ and amplitude $A/2\pi = 200\MHz$ (our method does not give any information about in-phase distortions).  
\item Construct $Y(f)$ by taking the discrete-time Fourier transform of the signal $y(t)$.
\item In this case, the input signal $x(t)$ is not an ideal delta or step function, but a Gaussian envelope with pulse width $\tpw=2.5\ns$ and amplitude $A/2\pi = 200\MHz$.  Create $x(t)$ by sampling the Gaussian at the same sample rate as the input $y(t)$. The complex part of $x(t)$ is zero.
\item Construct $X(f)$ by calculting the discrete-time Fourier transform of the signal $x(t)$.
\item Calculate the transfer function $\Ht(f) =Y(f)/X(f)$, as given by \EqRef{eq:H}.
\item Create the inverse $\Hi(f) = 1/\Ht(f)$.
\end{enumerate}
Once we have determined the transfer function and its inverse, we follow the same sequence as described in section S1 for creating the predistorted signal.

\end{widetext} 


\end{document}